\documentclass{raa}
\usepackage{graphicx,times}             %for PS/EPS graphics inclusion, new
\usepackage{natbib}
\usepackage{amssymb,amsmath}
\usepackage{mathrsfs}
\bibpunct{(}{)}{;}{a}{}{,}
\usepackage[a4paper=true,dvipdfm=true,pagebackref=true]{hyperref}
\hypersetup{colorlinks = true, linkcolor = red, anchorcolor = red, citecolor = blue, filecolor = red, pagecolor = pink, urlcolor = red}

\begin{document}

   \title{A correction method for the telluric absorptions and application to Lijiang Observatory}
   
   \volnopage{Vol.0 (20xx) No.0, 000--000}      %%preserved for Editor. DOn't remove!
   \setcounter{page}{1}          %%starting page, preserved for Editor. DOn't remove!

   \author{Kai-Xing Lu
      \inst{1,2}
   \and Zhi-Xiang Zhang
      \inst{3}
   \and Ying-Ke Huang
      \inst{4}
  \and An-Bing Ren
     \inst{5}
   \and Liang Xu
      \inst{1,2}
    \and Hai-Cheng Feng
      \inst{1,2}
   \and Yu-Xin Xin
      \inst{1,2} 
   \and Xu Ding
      \inst{1,2}
   \and Xiao-Guang Yu
      \inst{1,2}
    \and Jin-Ming Bai
      \inst{1,2}
     }

   \institute{Yunnan Observatories, Chinese Academy of Sciences, Kunming 650011, China; {\it lukx@ynao.ac.cn}\\
        \and
        Key Laboratory for the Structure and Evolution of Celestial Objects, Chinese Academy of Sciences, Kunming 650011, China\\
        \and
        Department of Astronomy, Xiamen University, Xiamen, Fujian 361005, China; {\it zhangzx@xmu.edu.cn}\\
        \and 
        Qian Xuesen Laboratory of Space Technology, China Academy of Space Technology, 104 Youyi Road, Haidian District, Beijing, 100094, China\\
        \and 
        Department of Astronomy, China West Normal University, 1 Shida Road, Shunqing District, Nanchong 637002, China\\
\vs\no
  % {\small Received~~20xx month day; accepted~~20xx~~month day}
  }

\abstract{ 
Observing a telluric standard star for correcting the telluric absorption lines of spectrum will take 
a significant amount of precious telescope time, especially in the long-term spectral monitoring project. 
Beyond that, it's difficult to select a  suitable telluric standard star near in both time and airmass to the scientific object.  
In this paper, we present a method of correcting the telluric absorption lines by combining the advantages of long slit spectroscopy. 
By rotating the slit, we observed the scientific object and a nearby comparison star in one exposure, so that 
the spectra of both objects should have the same telluric transmission spectrum. 
The telluric transmission spectrum was constructed by dividing the observed spectrum of comparison star by its stellar template, 
and was used to correct the telluric absorption lines of the scientific object. 
Using the long slit spectrograph of Lijiang 2.4-meter telescope, we designed a long-term spectroscopic observation strategy, 
and finished a four-year spectroscopic monitoring for a pair of objects (an active galactic nuclei and an non-varying comparison star). 
We applied this method to correct the telluric absorption lines of the long-term monitored spectra by Lijiang 2.4-meter telescope, 
and investigated the variation of the telluric absorptions at Lijiang Observatory. 
We found that the telluric absorption transparency 
is mainly modulated by the seasonal variability of the relative humidity, airmass and seeing. 
Using the scatter of the [O~{\sc III}]~$\lambda$5007 fluxes emitted from the narrow-line region of active galactic nuclei as an indicator, 
we found that the correction accuracy of the telluric absorption lines is 1\%. 
}

   \authorrunning{K.-X. Lu et al. }            %author_head in even pages
   \titlerunning{Correcting telluric absorptions and application}  % title_head in odd pages

   \maketitle

\section{Introduction}           %% first-level sections will be auto-capitalized
\label{sect:intro}
Ground-based optical and near-infrared spectra are contaminated 
by the variable absorptions of the telluric atmosphere. In order to retrieve the intrinsic 
spectral signal of the scientific object, the traditional method is to observe a telluric standard star (early-type star) near in both 
time and airmass to the scientific object, 
then using the telluric spectrum created from standard star to correct the telluric absorption features by dividing the object's spectrum during the data reduction. 
Often, a rapidly rotating hot star (spectral type A or late B) is frequently adopted as the telluric standard 
because there are few and weak metal lines in the spectrum around the telluric absorption bands. 

In addition, many sophisticated methods have been developed. For example, using G2 V star as telluric standards and the 
high-resolution spectrum of the sun,  \cite{Maiolino1996} developed a correction method for the telluric absorption lines. 
Combining \cite{Maiolino1996}'s method with above traditional early-type star method, \cite{Hanson1996,Hanson2005} 
developed a method to correct the telluric absorption lines, which has to observe an A-star and a G2 V star 
and involves a somewhat tedious and complicated reduction process. 
\cite{Vacca2003} described a new method of generating a telluric correction spectrum by using 
observations of A0 V stars and a high-resolution model spectrum of Vega, which is similar to the process described 
by \cite{Maiolino1996}. 

Another approach is to generate a theoretical telluric absorption spectrum of the atmospheric transmission from a 
line database and the observing condition. 
Based on the available radiative transfer model, this approach has been successfully developed  by several groups 
(e.g., \citealt{Seifahrt2010,Gullikson2014,Bertaux2014,Cotton2014}). 
Furthermore, some groups incorporated the radiative transfer model into they own code to fit an entire input 
spectrum (e.g., \citealt{Husser2013,Gullikson2014,Smette2015}). 

Actually, above methods have several disadvantages(e.g., \citealt{Seifahrt2010,Gullikson2014}): 
(1) it's difficult to find a suitable telluric standard star near in both time and airmass to the scientific object during the observation. 
(2) The strength of telluric absorption lines may significantly changes with the varying observing conditions (such as the weather, airmass and seeing), 
even if we find an ideal telluric standard star, 
it is almost impossible to obtain the spectrum of scientific object and telluric standard star under the same observing conditions.  
(3) if we conduct a long-term spectral monitoring study, observing a telluric standard star near the same time to the object, 
we will take a significant amount of precious telescope time. 
(4) The deviation of the modelled telluric absorption spectrum is mainly due to 
the limited accuracy of the meteorological data and errors in the radiative transfer model. 
Therefore, in this paper, we described a new method of correcting the telluric absorption lines from the spectrum, 
and used it to investigate the variation of telluric absorptions of Lijiang Observatory. 

The paper is organised as follows. In Section~\ref{sec:method}, 
we described the correction method for the telluric absorption lines of spectrum. 
We applied this method to the spectra observed by Lijiang 2.4-meter telescope 
in Section~\ref{sec:app}, including observation, data reduction, the telluric absorption correction, 
accuracy evaluation, and investigated the telluric absorption features of Lijiang Observatory. 
Section~\ref{sec:dis}~and~\ref{sec:sum} are discussion and summary, respectively. 

\section{Method}
\label{sec:method}

\subsection{Long slit spectroscopy}
The spectrograph used in astronomy often equipped with long slit (with different projected slit width), 
so that we can orient long slit to 
take the spectra of the scientific object and a nearby non-varying comparison star simultaneously. 
This observation method is widely used in the reverberation mapping of active galactic nuclei to measure 
supermassive black hole mass and investigate the physics of the broad-line region 
(e.g., \citealt{Maoz1990,Kaspi2000, Lu2019}), because the spectrum of the non-varying comparison star 
can be used as the fiducial spectrum to calibrate the flux of scientific object in high precision (e.g., see \citealt{Lu2019}). 
The second application of the comparison star spectrum is to correct the line broadening caused by seeing and instruments \citep{Du2016}. 
In the application of correcting instrument broadening, the line broadening function can be obtained from the fitting between 
the comparison star spectrum and its stellar template, then we use the broadening function to correct the broadening of 
the emission-line profile through Richardson$-$Lucy deconvolution algorithm \citep{Richardson1972,Lucy1974}. 
Here we provide a new application of the comparison star spectrum. 

\subsection{Fitting spectrum and correcting telluric absorptions} 
\begin{figure*}
   \centering
   \includegraphics[width=10cm, angle=0]{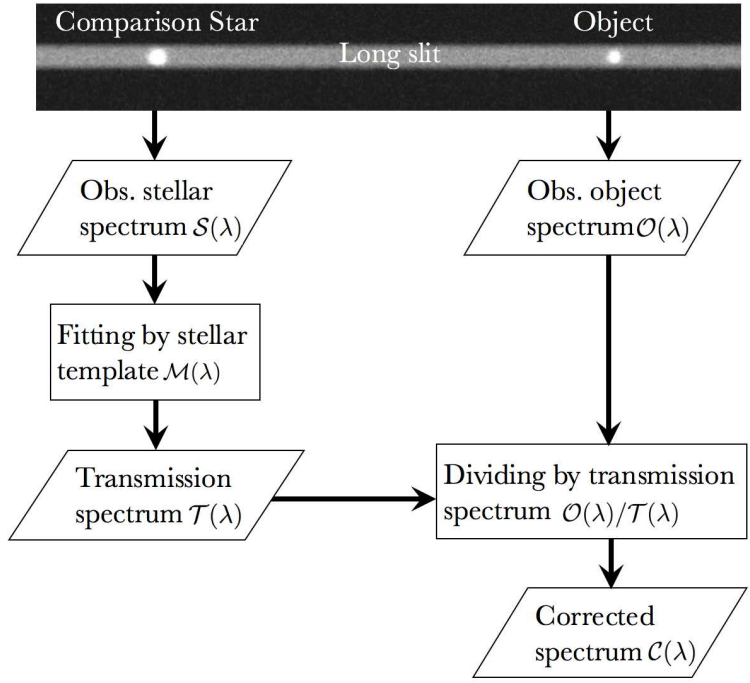}
   \caption{Schematic diagram of correction for the telluric absorption lines.}
   \label{Fig:sd}
 \end{figure*}

In the scheme of long slit spectroscopy, the spectra of the scientific object and comparison star are obtained 
simultaneously. Extraction processes of the observed object and stellar spectra are similar to 
the works of \cite{Lu2016} and \cite{Lu2019} (more details refer to Section~\ref{sec:app}). 
The schematic diagram of correcting the telluric absorption lines shows in Figure~\ref{Fig:sd}. 
We first obtain the telluric transmission spectrum by fitting the spectrum of 
comparison star by a stellar template. Generally, there are two approaches to select the 
stellar template of comparison star: 
(1) selecting the most similar stellar spectrum from the library of synthesis stellar model as the 
template of comparison star (e.g., \citealt{Husser2013}). 
(2) searching the calibrated spectra of comparison star from spectral survey database 
(e.g., SDSS, LAMOST) as the template of comparison star\footnote{https://github.com/BU-hammerTeam/PyHammer}. 
To fit the observed spectrum of comparison star, 
the selected stellar template ${\mathcal M}(\lambda)$ is shifted, broadened and scaled. 
That is the observed spectrum ${\mathcal S}(\lambda)$ of comparison star is modelled by function 
\begin{equation}
  {\mathcal S'}(\lambda)={\mathcal P}(\lambda)[{\mathcal M}(\lambda) \otimes {\mathcal G}({\rm Shift,FWHM})]. 
 \label{eq:model}
\end{equation}
Where,  ${\mathcal P}$ is a Legendre polynomial that accounts for the mismatch in the continuum shape of 
comparison star and stellar template, ${\mathcal G}$ is a Gaussian kernel with parameters of velocity and shift 
that accounts for the mismatch in the line width and shift of the intrinsic absorption lines. 

It's should be noted that the telluric absorptions from the red side of the optical spectrum to the near infrared 
are mainly caused by $\rm O_{2}$ and $\rm H_{2}O$. The main telluric absorption bands include narrow 
oxygen bands (identified as the Fraunhofer $\gamma$, B, and A bands) and some comparably weak water vapour features. 
The telluric absorption bands of optical region with bandwidths (i.e., absorption windows) are presented in Table~\ref{Tab} 
(see also \citealt{Smette2015}). During the fitting of the observed spectrum, 
the spectral region within the telluric absorption windows (see Table~\ref{Tab}) are masked. 
Then the telluric transmission spectrum ${\mathcal T}(\lambda)$ in the selected absorption windows is constructed 
by dividing the observed spectrum of comparison star ${\mathcal S}(\lambda)$ by the modelled spectrum ${\mathcal S'}(\lambda)$. 
Lastly, we obtained the telluric absorption corrected spectrum by dividing the observed object 
spectrum ${\mathcal O}(\lambda)$ by the telluric transmission spectrum ${\mathcal T}(\lambda)$, 
because the spectra of comparison star and the scientific object are obtained at the same time and with the same airmass and seeing. 
In fact, this step resembles the process of {\tt IRAF} telluric. 
Next, we use this method to correct the telluric absorption lines of the spectra observed by Lijiang 2.4-meter telescope 
and investigate the variation of the telluric absorptions at Lijiang Observatory. 

\begin{table}
\begin{center}
\caption[]{Telluric absorption bands and selected windows in the optical region. ``${\rm B*}$'' means the number 
of the telluric absorption bands (Col. 1), these absorption bands with different absorption widths (Col. 3) are caused 
by different absorbers (Col. 2).}\label{Tab} 
 \begin{tabular}{ccc}
  \hline\noalign{\smallskip}
No  &  Molecular &  Windows                      \\
(1)&(2)&(3)\\
  \hline\noalign{\smallskip}
B1  & $\rm O_{2}~(\gamma~band)$         & 6270~\AA$\sim$6350~\AA \\
B2  & $\rm O_{2}~(B~band) + H_{2}O$    & 6830~\AA$\sim$7050~\AA \\
B3  & $\rm H_{2}O$                                  & 7143~\AA$\sim$7369~\AA \\
B4  & $\rm O_{2}~(A~band)$                    & 7550~\AA$\sim$7750~\AA \\
B5  & $\rm H_{2}O$                                  & 8080~\AA$\sim$8380~\AA \\
  \noalign{\smallskip}\hline
\end{tabular}
\end{center}
\end{table}

\begin{figure}
   \centering
   \includegraphics[width=\textwidth, angle=0]{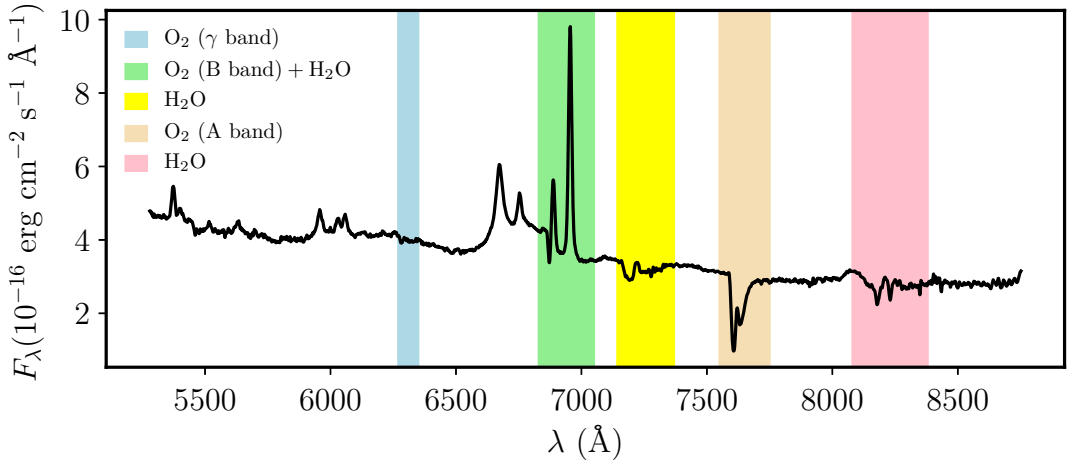} 
   \caption{The spectrum of SDSS J153636.22+044127.0 (J1536) observed by Lijiang 2.4-meter telescope. 
   Five telluric absorption bands in the observed spectrum are marked by colour diagram (see Table~\ref{Tab}).} 
   \label{Fig:obs}
\end{figure}

\section{Application}
\label{sec:app}
The Lijiang 2.4-meter Telescope (LJT) is located at Lijiang Observatory 
(E: $100^{\circ} 01'48''$, N:$26^{\circ}41'42''$, altitude: 3193~meters), 
and is administered by Yunnan Observatories of the Chinese Academy of Sciences. 
LJT observation time is opening to the worldwide astronomical community since 2008. 
The structure of the 2.4-meter telescope including optical, mechanical, motion and control systems 
along with its instruments are described by \cite{Fan2015} and \cite{Wang2019}. Basic observing conditions 
(including seeing, wind speed, weather and so on) are presented in recent paper of \cite{Xin2020}. 
The Yunnan Faint Object Spectrograph and Camera (YFOSC) mounted on the 2.4-meter 
telescope is a versatile instrument for spectroscopy and photometry. 
As a versatile instrument, YFOSC is equipped with a series of Grisms, long slits and filters. 
The wavelength coverage of these Grisms is about 3600~\AA $\sim$9000~\AA, the red side of 
spectrum observed by YFOSC is covered by some telluric absorption bands.  
Correction of the telluric absorption lines is an essential step to extract the spectral signals. 

\subsection{Observation strategy and spectral monitoring}
\label{sec:obs}
To estimate the correction accuracy of the telluric absorption lines using above described method, 
and to investigate the variation of the telluric absorptions at Lijiang observatory, we designed a 
strategy of long slit spectroscopy and developed a long-term spectroscopic monitoring project using Lijiang 2.4-meter telescope. 
In the observation of active galactic nuclei, it is generally believed that the fluxes of forbidden line [O~{\sc III}]~$\lambda$5007 
emitted from the narrow-line region are stable in the timescale of years (e.g., \citealt{Foltz1981,Lu2016,Lu2019}). 
If the wavelength of [O~{\sc III}]~$\lambda$5007 emission line locate in the telluric absorption band in the observed frame, the apparent variation of 
the [O~{\sc III}]~$\lambda$5007 fluxes will be an indicator of the telluric absorptions and the scatter of [O~{\sc III}]~$\lambda$5007 fluxes 
after correcting the telluric absorptions can also be used to estimate the correction accuracy. 
Therefore, we orient the long slit to take the spectra of the object SDSS J153636.22+044127.0 
(J1536) and an non-varying comparison star simultaneously, where J1536 is an active galactic nuclei (AGN) with redshift 0.3889, 
its optical spectrum observed by Lijiang 2.4-meter telescope showed in Figure~\ref{Fig:obs}. 
There are five telluric absorption bands in the observed spectra, which are listed in Table~\ref{Tab} and marked in Figure~\ref{Fig:obs}. 
In the observed frame, the [O~{\sc III}]~$\lambda$5007 emission line of J1536 is in the 
B2 absorption band (see Figure~\ref{Fig:obs} and Table~\ref{Tab}), 
we can measure the scatter of the [O~{\sc III}]~$\lambda$5007 fluxes before and after correcting the telluric absorption lines 
and use to estimate the correction accuracy of the telluric absorption lines. 
On the other hand, J1536 is a candidate of supermassive black hole binary (SMBHB, e.g., \citealt{Boroson2009,Lu2016b,Zhang2019}), 
except for investigating the variation of the telluric absorptions, we will use these long-term monitored spectra to investigate 
the broad-line region and optical variability physics of SMBHB. The later will be presented in a separation paper. 

We have finished a four-year spectroscopic monitoring for a pair of objects (J1536 $vs.$ its comparison star). 
The observation started from 2017 and continued to 2020, 
and the annual spectral sampling lasts from January to September. 
All spectra of J1536 and comparison star are observed using Grism 8 mounted on the terminal of YFOSC, 
which provides a resolution of 78~\AA~mm$^{-1}$ (1.5~\AA~pixel$^{-1}$) and covers the wavelength range 5000~\AA $\sim$9000~\AA, 
which fully guaranteed the study of long-term variation of the telluric absorptions at Lijiang Observatory. 

 \begin{figure}
   \centering
   \includegraphics[width=\textwidth, angle=0]{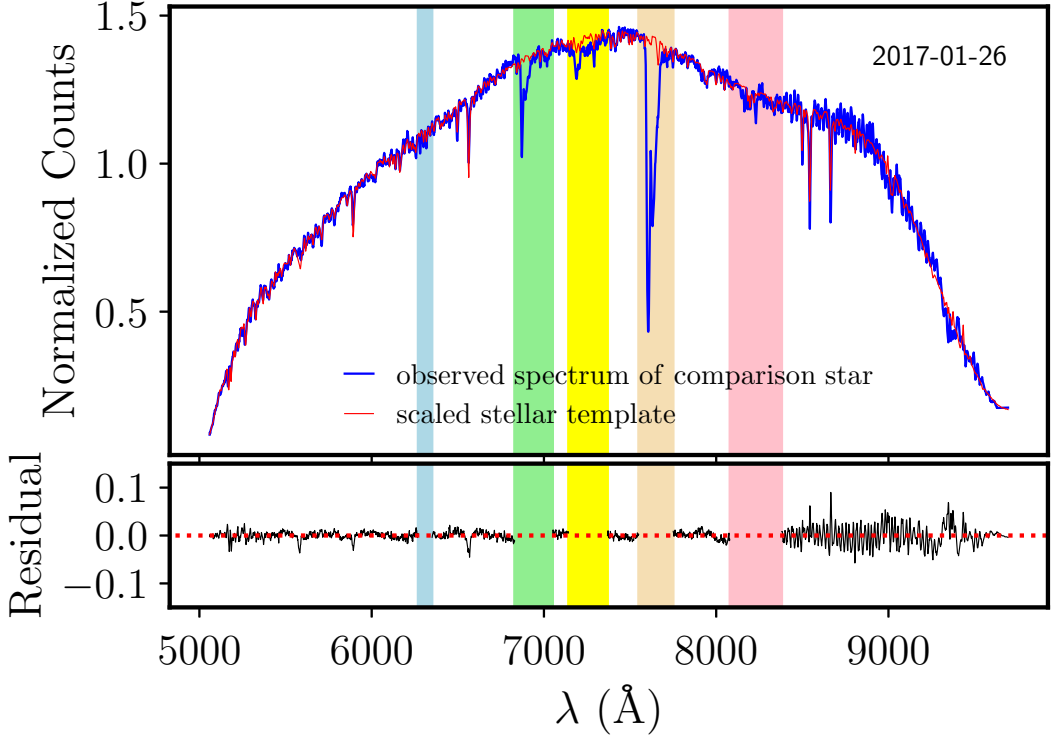}
   \caption{The top panel is an example of fitting observed spectrum ${\mathcal C}(\lambda)$ by the stellar template, the observed spectrum of comparison star in blue, and scaled stellar template ${\mathcal C'}(\lambda)$ in red.  
   The bottom panel is the residual of fitting regions.}
   \label{Fig:fit}
 \end{figure}

\subsection{Data reduction}
\label{sec:dr}
Following the standard process included bias subtraction, flat-field correction, wavelength calibration and spectrum extraction, 
we reduced the two-dimensional spectroscopic data using the standard {\tt IRAF(v2.16)} package. 
All spectra were extracted using a uniform aperture of 20 pixels (5.7$^{\prime\prime}$), 
and background was determined from two adjacent regions ($+7.4^{\prime\prime}\sim+14^{\prime\prime}$ 
and $-7.4^{\prime\prime}\sim-14^{\prime\prime}$) on both sides of the aperture region. 
Spectral fluxes of the scientific object were calibrated by the comparison stars in two steps. 
(1) We produced a fiducial spectrum of the comparison star using the data from nights with photometric conditions. 
(2) For each object/comparison star pair, we obtained a wavelength-dependent sensitivity function by comparing 
the star's spectrum to the fiducial spectrum. 
Then this sensitivity function was applied to calibrate the observed spectrum of the scientific object (also see \citealt{Lu2016,Lu2019}). 

\subsection{Correction of the telluric absorptions} 
\label{sec:c}
\begin{figure}
   \centering
   \includegraphics[width=0.49\textwidth, angle=0]{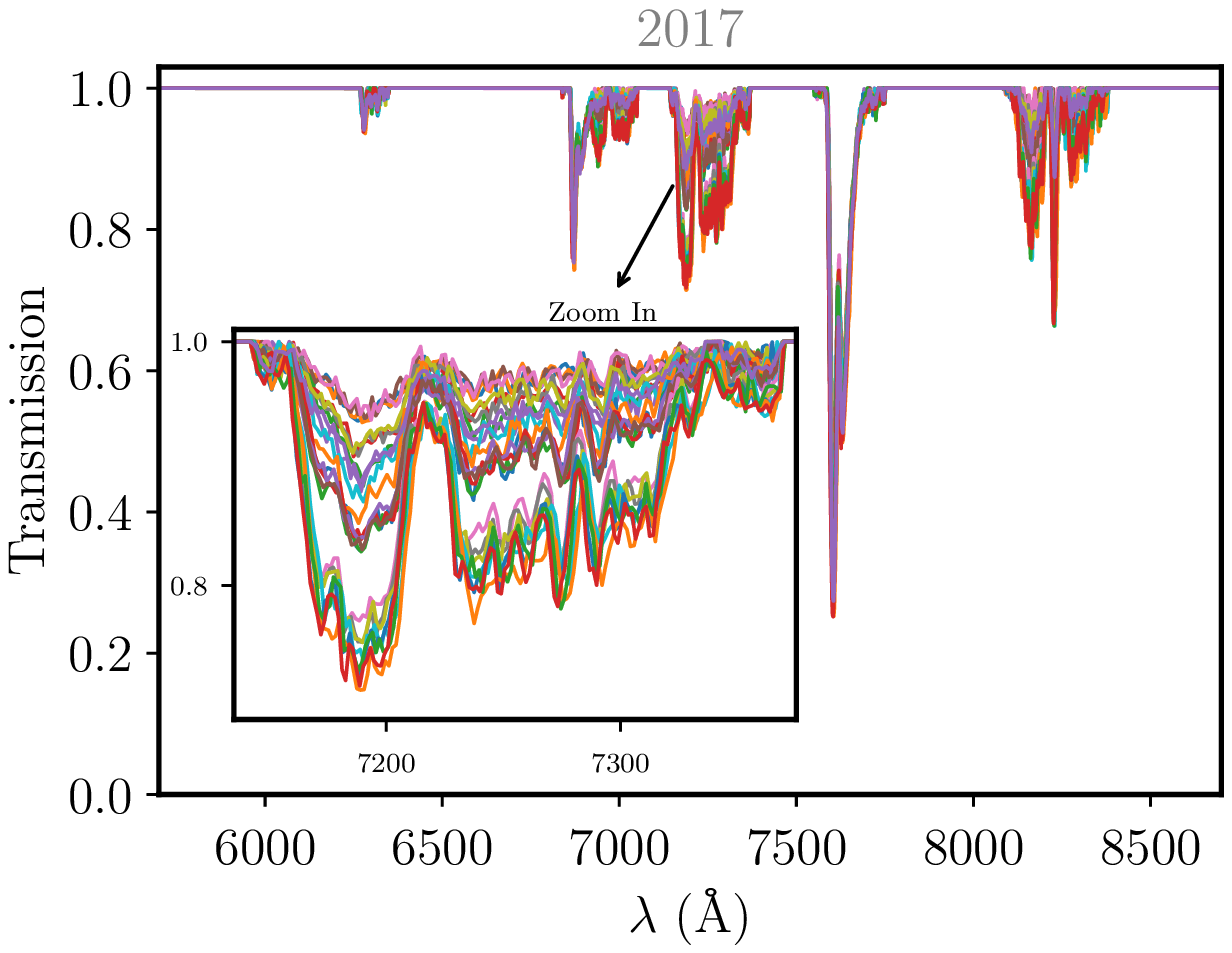}
   \includegraphics[width=0.49\textwidth, angle=0]{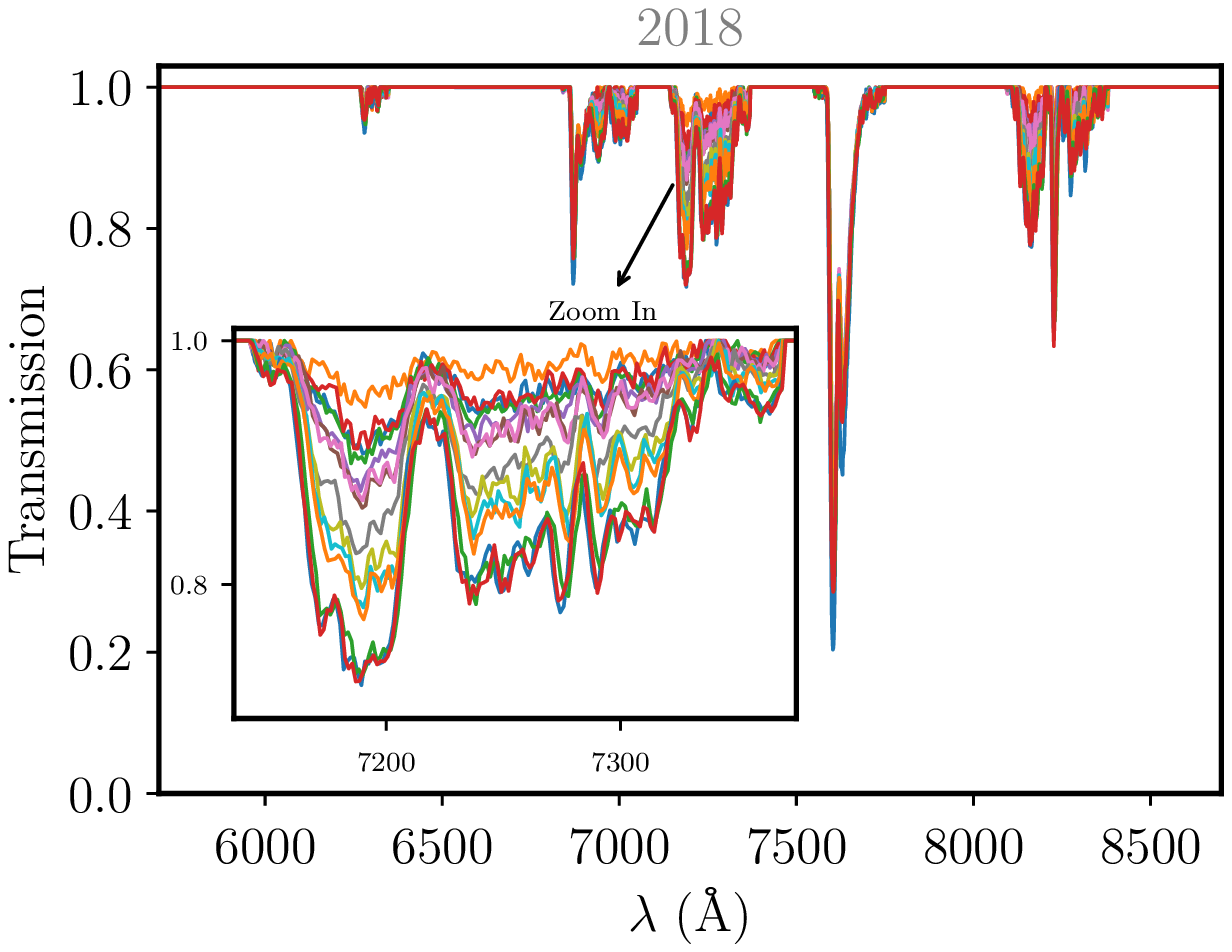}\\
   \includegraphics[width=0.49\textwidth, angle=0]{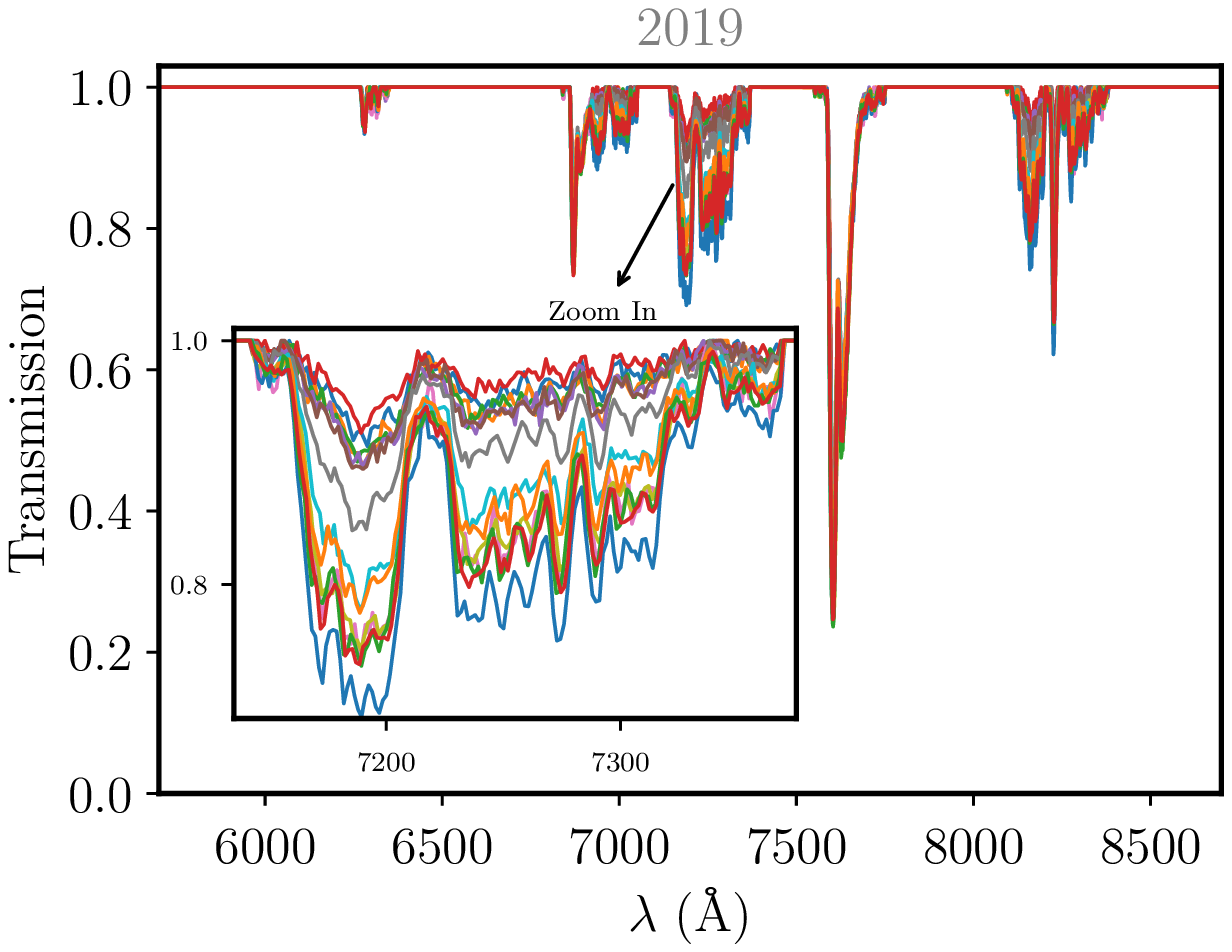}
   \includegraphics[width=0.49\textwidth, angle=0]{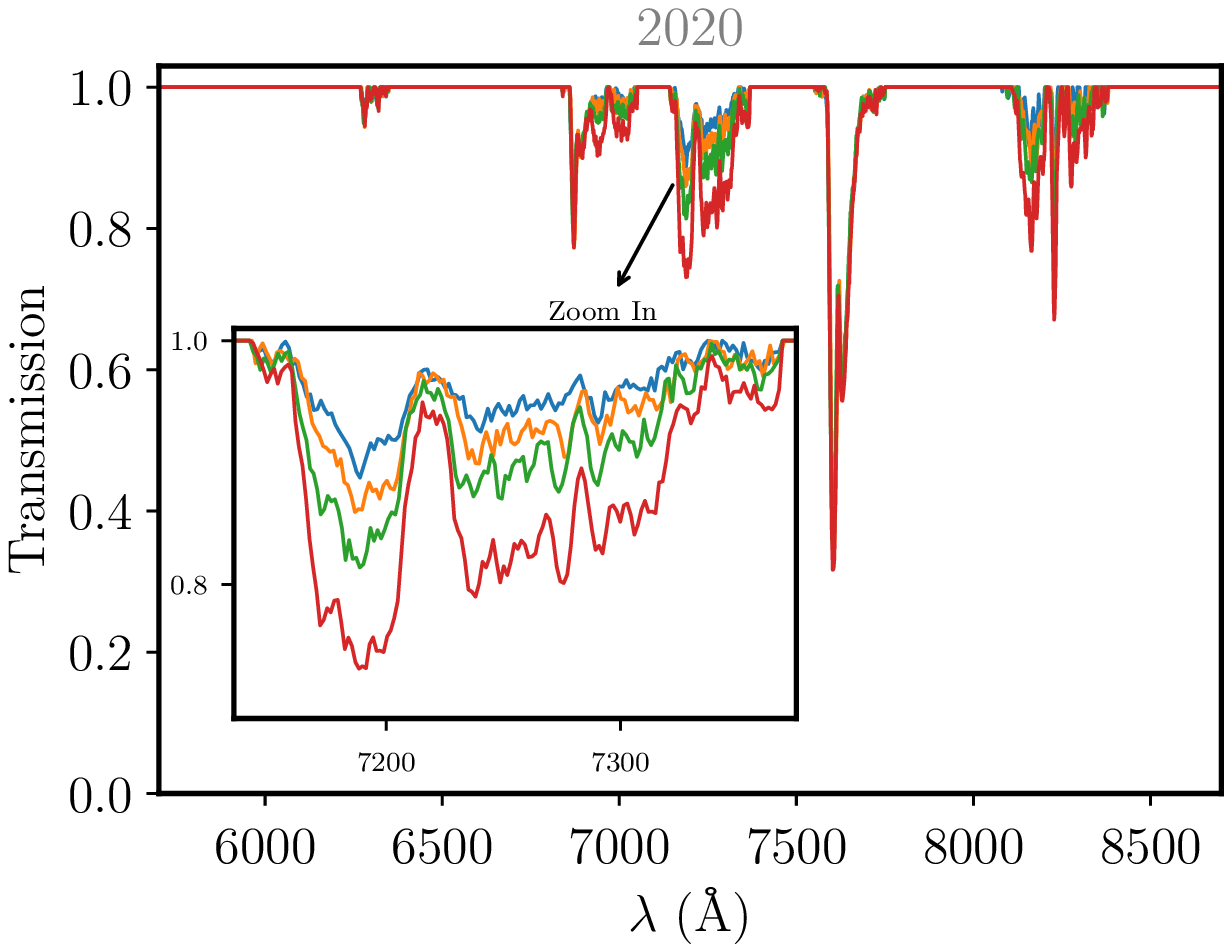}
   \caption{The transmission spectra of Lijiang Observatory obtained in Section~\ref{sec:c} provide values from 0 (totally opaque) to 1 (fully transparency). 
   The year of observation was noted in the title of panel, a transmission spectrum corresponds to one observation.}
   \label{Fig:trans}
  \end{figure}

\begin{figure}
   \centering
   \includegraphics[width=\textwidth, angle=0]{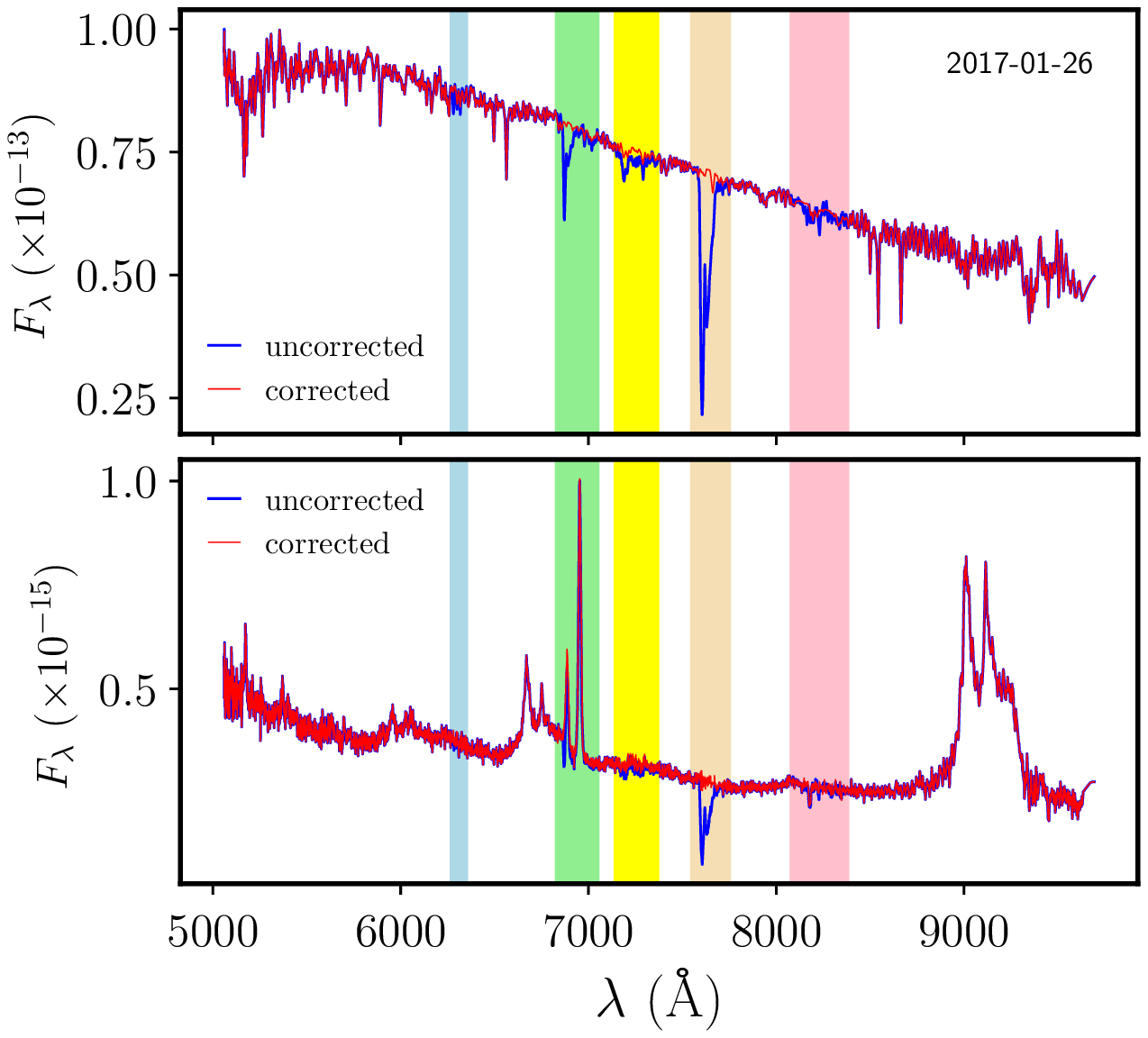}
   \caption{Comparison of the telluric absorption corrected spectra (see Section~\ref{sec:c}) and uncorrected spectra in the observed frame. 
   Only flux calibrated spectra show in blue, 
   the telluric absorption corrected and flux calibrated spectra show in red. The units of $F_{\lambda}$ is erg~s$^{-1}$~cm$^{-2}$~\AA$^{-1}$. 
   The top panel shows the spectra of comparison star, 
   and the bottom panel shows the spectra of J1536.} 
   \label{Fig:obs_cali}
  \end{figure}

In practice, we would better correct the telluric absorption at first, and then calibrate spectral fluxes. 
In this case, we do not need to mask the telluric absorption bands again during the spectral flux calibration. 
Following the method described in Section~\ref{sec:method}, we first use the Equation~\ref{eq:model} to model the observed 
spectra ${\mathcal S}(\lambda)$ of comparison star (in counts). 
Because the wavelength coverage of all observed spectra is about 
5000~\AA $\sim$9000~\AA, these significant absorption bands listed in Table~\ref{Tab} are included, 
and they are masked in modelling process (see Figure~\ref{Fig:fit}). Then we obtained the telluric transmission spectrum by dividing the observed 
spectrum ${\mathcal S}(\lambda)$ of comparison star by the modelled spectrum ${\mathcal S'}(\lambda)$. In order to avoid introducing new errors 
to the corrected spectrum of the scientific object, we set the transmission of the wavelength regions outside the telluric absorption windows to 1 
before correcting the telluric absorption lines of the observed spectrum. This is reasonable 
because $\rm O_{2}$ and $\rm H_{2}O$ molecular are fully transparent for the photons from these wavelength regions (see \citealt{Smette2015}). 
All transmission spectra (in the telluric absorption windows) calculated from above process were plotted annually in Figure~\ref{Fig:trans}, 
we zoom in the B3 absorption bands and over-plotted in Figure~\ref{Fig:trans}. 
The figure clearly shows that the telluric absorption depth presents significant variation. 
We first used the telluric transmission spectra to correct the telluric absorption lines of the observed spectra, 
then we obtained flux calibrated spectra using the steps described in Section~\ref{sec:dr}. 
Figure~\ref{Fig:obs_cali} is an example for comparing the corrected spectra of telluric absorptions and uncorrected spectra, 
where the top panel is the spectra of comparison star, and the bottom panel is the spectra of J1536. 
  
\subsection{Correction accuracy of the telluric absorptions} 
As stressed in Section \ref{sec:obs}, we selected a special AGN of J1536 with redshift 0.3889 to put its 
[O~{\sc III}]~$\lambda$5007 emission line drop into B2 telluric absorption bands, and developed a long-term 
spectroscopic monitoring, so that we can use the scatter of the [O~{\sc III}]~$\lambda$5007 fluxes to estimate 
the correction accuracy of the telluric absorption lines. Strictly speaking, the [O~{\sc III}]~$\lambda$5007 emission line 
of J1536 drops into the red side of B2 telluric absorption bands that dominated by H$_{2}$O molecular absorption. 

In practice, the fluxes of forbidden line [O~{\sc III}]~$\lambda$5007 emitted from the narrow-line region of active galactic nuclei 
is a constant over reverberation timescales ($\sim$ years), therefore it can be used as an internal flux calibration 
standard (e.g, \citealt{Foltz1981,Peterson1982,Peterson2002}). 
Based on reverberation mapping of active galactic nuclei undertook by Lijiang 2.4-meter telescope, 
\cite{Lu2016} found that the [O~{\sc III}]~$\lambda$5007 fluxes of NGC~5548 measured from flux calibrated spectra show 
a small fluctuation at a level of 2\%, which is regarded as the accuracy of the flux calibration. 
While, \cite{Lu2019} recently found that the [O~{\sc III}]~$\lambda$5007 fluxes of Mrk~79 give rise to 
the apparent variation at a level of 5\% over reverberation timescales. \cite{Lu2019} further found that, 
if the [O~{\sc III}]~$\lambda$5007 emission region is an extended source, 
the varying observing conditions (e.g., varying seeing) would cause extra dispersion of the [O~{\sc III}]~$\lambda$5007 fluxes. 
It should be noted that the [O~{\sc III}]~$\lambda$5007 
emission region of NGC~5548 approximates to point source, and the [O~{\sc III}]~$\lambda$5007 emission region 
of Mrk~79 is an extended source. 

\begin{figure}
   \centering
   \includegraphics[width=\textwidth, angle=0]{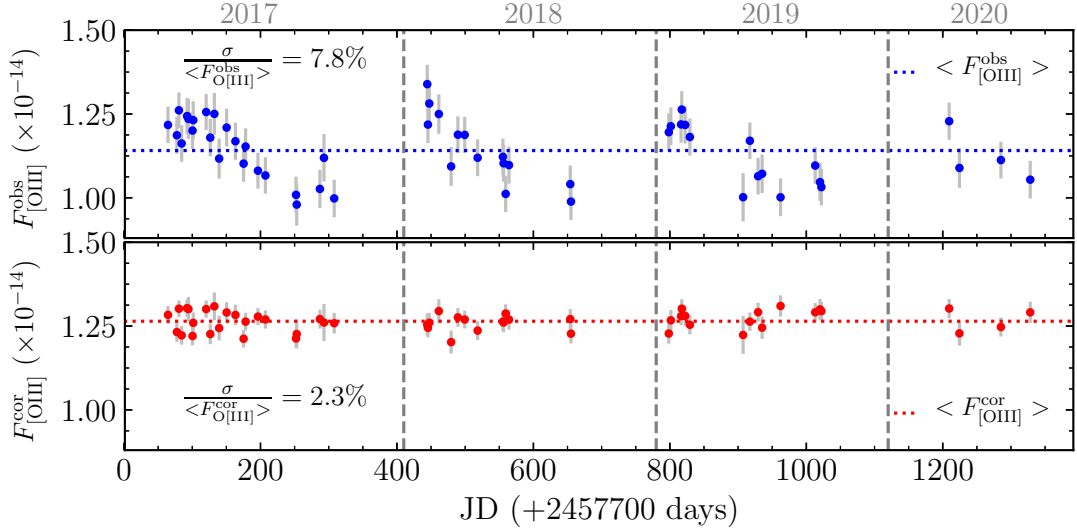}
   \caption{The [O~{\sc III}]~$\lambda$5007 fluxes of J1536 in units of erg~s$^{-1}$~cm$^{-2}$. The top panel shows the apparent variation 
   of the [O~{\sc III}]~$\lambda$5007 fluxes measured from the flux calibrated spectrum (before correcting the telluric absorption lines), 
   periodicity is caused by the seasonal variability of the relative humidity. 
   The bottom panel shows the scatter of the [O~{\sc III}]~$\lambda$5007 fluxes measured from the flux calibrated and 
   the telluric absorption corrected spectra. The year of observation was noted in the title of panel.}
   \label{Fig:oiii}
  \end{figure}

To directly understand the impacts of the telluric absorptions on the scientific spectra, 
we measured the [O~{\sc III}]~$\lambda$5007 fluxes of J1536 from the flux calibrated spectra.  
We used two continuum bands (4710$-$4740~\AA~and 5080$-$5110~\AA~in the rest frame) 
to set the continuum underneath the [O~{\sc III}]~$\lambda$5007 and H$\beta$ emission lines, 
then we integrated the continuum-subtracted [O~{\sc III}]~$\lambda$5007 fluxes between 4980~\AA~and 5025~\AA, 
and presented in the top panel of Figure~\ref{Fig:oiii}. 
The figure shows the [O~{\sc III}]~$\lambda$5007 fluxes 
before correcting the telluric absorption lines demonstrate apparent period variation with a scatter of 7.8\%.  
This scatter is larger than previous observation record (such as 5\% of Mrk~79 and 2\% of NGC~5548). 
The apparent period variation follows the seasonal variability of the relative humidity of Lijiang Observatory 
(see Figure 22 of \citealt{Xin2020}). Qualitatively, January is the deep winter of Lijiang Observatory, 
the atmospheric relative humidity is the lowest, then begins to increase slowly, and reaches the maximum 
in the rainy season (about from July to September ). 
Therefore,  accurate calibration of the telluric absorption lines from observed spectrum is crucial for 
obtaining credible scientific data. 

\begin{figure}
   \centering
   \includegraphics[width=0.8\textwidth, angle=0]{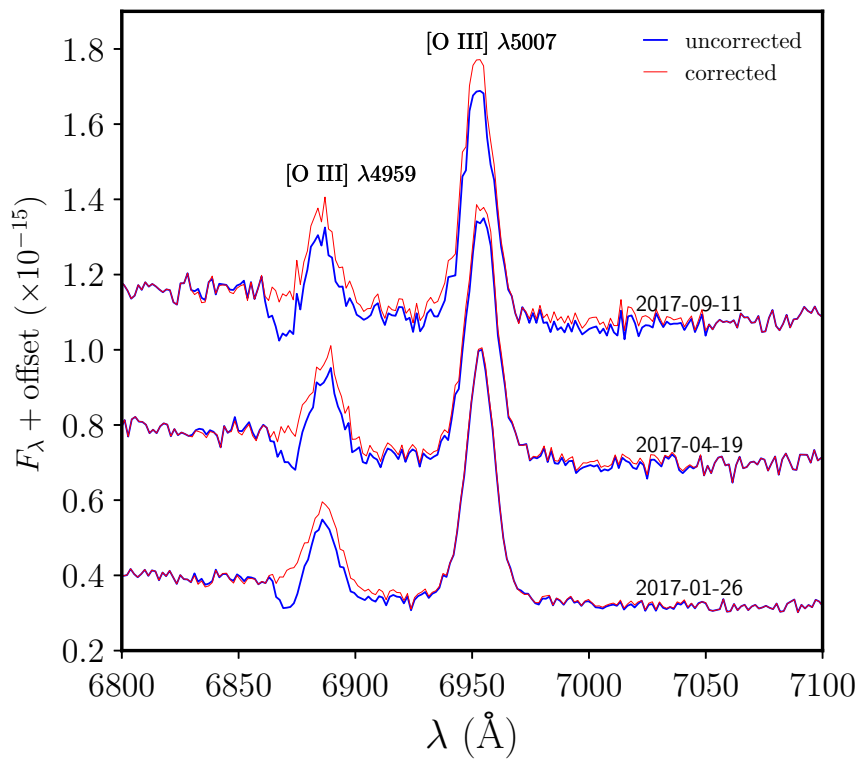}
   \caption{The telluric absorption corrected and uncorrected [O~{\sc III}] profiles from the observation of 2017 (in the observed frame), 
   observation dates are noted, the units of $F_{\lambda}$ is erg~s$^{-1}$~cm$^{-2}$~\AA$^{-1}$. 
   The differences of [O~{\sc III}] profiles before and after correcting the telluric absorptions are presented clearly.} 
   \label{Fig:oiiireg}
  \end{figure}

As a comparison, we also measured the [O~{\sc III}]~$\lambda$5007 fluxes of J1536 from the flux calibrated and the telluric absorption corrected spectra. 
The measurement results were plotted in the bottom panel of Figure~\ref{Fig:oiii} for comparing with the measurements 
before correcting the telluric absorption lines, which shows that the scatter of the [O~{\sc III}]~$\lambda$5007 
fluxes is 2.3\%, and the apparent period variation presented in the top panel is disappeared. 
As addressed above, the fluxes of [O~{\sc III}]~$\lambda$5007 are stable enough in the timescale of years 
if the [O~{\sc III}]~$\lambda$5007 emission region is compact. In practice, the scatter of [O~{\sc III}]~$\lambda$5007 fluxes (2.3\%) 
is caused by the accuracy of flux calibration and the accuracy of the telluric absorption  correction. 
If we assume that the flux calibration accuracy of J1536 is the same with our previous experiment of NGC~5548 (i.e., 2\%, \citealt{Lu2016}), 
this means that the correction accuracy of the telluric absorption lines using the method described in Section~\ref{sec:method} is 1\%. 
In addition, in order to intuitively inspect the differences before and after correcting the telluric absorptions, 
we plotted the telluric absorption corrected and uncorrected [O~{\sc III}] profiles selected randomly from observations of 2017 in Figure~\ref{Fig:oiiireg}. 
The figure also demonstrates the differences of the telluric absorption from the different observations. 

 \begin{figure}
   \centering
   \includegraphics[width=\textwidth, angle=0]{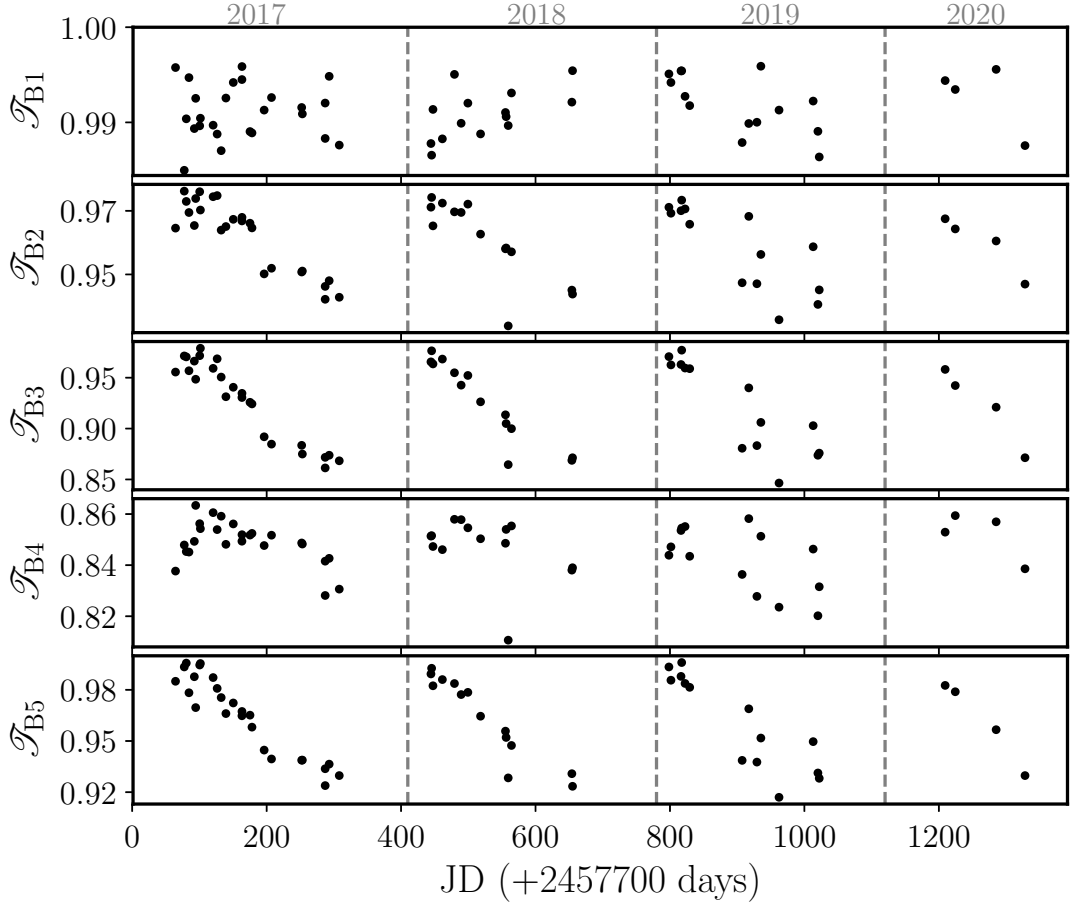}
   \caption{The integral transparency ($\mathscr{T}_{\rm B1}$, $\mathscr{T}_{\rm B2}$, $\mathscr{T}_{\rm B3}$, $\mathscr{T}_{\rm B4}$ 
   and $\mathscr{T}_{\rm B5}$) of the telluric absorption band (B1, B2, B3, B4 and B5, see Table~\ref{Tab}) as a function of time, respectively. 
   The observation started from 2017 and continued to 2020, and the annual spectral sampling almost lasts from January to September. 
   The year of observation was noted in the title of panel.}
   \label{Fig:transp}
  \end{figure}

 \begin{figure}
   \centering
   \includegraphics[width=\textwidth, angle=0]{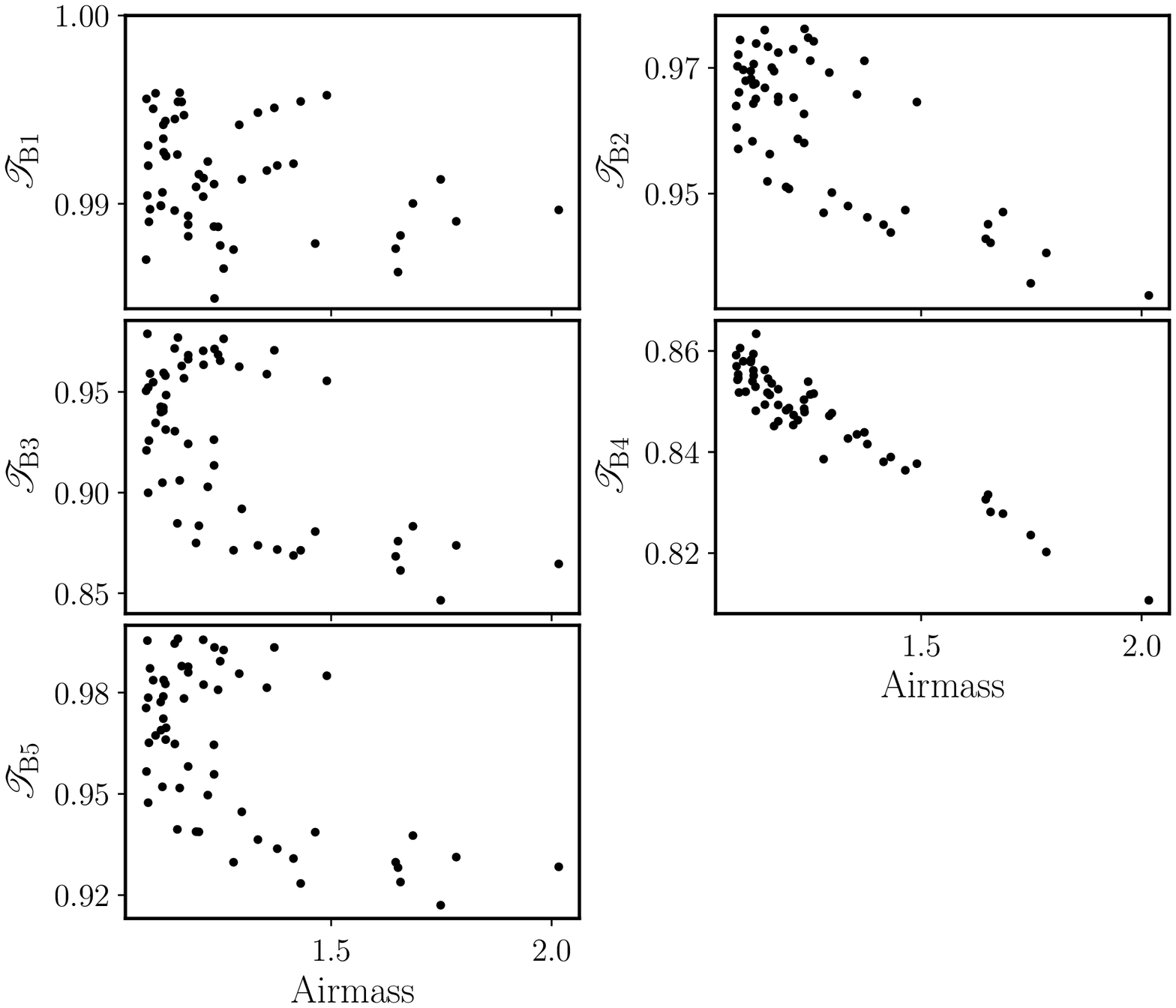}
   \caption{Relationship between the integral transparency and airmass. The results of correlation analysis are listed in Table~\ref{Tab:sta}.}
   \label{Fig:relation}
  \end{figure}

\subsection{Telluric absorption features of Lijiang Observatory}
\label{sec:lo}
In order to understand the variation of the telluric absorptions caused by the molecular of 
$\rm O_{2}$, $\rm H_{2}O$ or both together at Lijiang Observatory, 
we calculated the integral transparency (denoted by $\mathscr{T}_{\rm B*}$) of each absorption band using formula 
of $\int_{\lambda1}^{\lambda2} {\mathcal T}(\lambda)\,d\lambda/(\lambda2-\lambda1)$ 
(where $\lambda1$ and $\lambda2$ are the edges of the telluric absorption bands, ${\mathcal T}(\lambda)$ is the telluric transmission spectrum), 
and plotted them as the functions of time in Figure~\ref{Fig:transp}. Where ``${\rm B*}$'' means the number 
of the telluric absorption band, which corresponds to the number of Table~\ref{Tab}. 
Inspecting the figure, we can see that the variation of the integral transparency for the telluric absorption bands mainly or fully caused by 
the molecular of $\rm H_{2}O$ (i.e., B2, B3 and B5 absorption bands) is similar with the apparent period variation 
of the [O~{\sc III}]~$\lambda$5007 fluxes that contaminated by the $\rm H_{2}O$ absorption (the top panel of Figure~\ref{Fig:oiii}). 
The seasonal variability of the $\rm H_{2}O$ absorption bands is clearly visible, 
which is consistent with the seasonal variability of the relative humidity (see also Figure 22 of \citealt{Xin2020}). 
While, the integral transparency for the absorption bands caused by the molecular of $\rm O_{2}$ (i.e., B1 and B4 absorption bands)
do not present obvious variation trend in the time domain. 
In the varing observing conditions, the averaged integral transparency
($\widehat{\mathscr{T}_{\rm B1}}$, $\widehat{\mathscr{T}_{\rm B2}}$, $\widehat{\mathscr{T}_{\rm B3}}$, $\widehat{\mathscr{T}_{\rm B4}}$ and $\widehat{\mathscr{T}_{\rm B5}}$) of the telluric absorption bands (B1, B2, B3, B4 and B5) is (0.991, 0.961, 0.926, 0.847 and 0.964) 
with scatter (i.e., standard deviation) of (0.3\%, 1.1\%, 3.9\%, 1.1\% and 2.4\%), 
this means that about (1\%, 4\%, 7\%, 15\% and 4\%) of photons propagated from the distant 
objects are absorbed by the $\rm O_{2}$ or $\rm H_{2}O$ in these absorption windows, respectively. We also listed the minimum and maximum 
of the integral transparencies in Table~\ref{Tab:sta}. 

\begin{figure}
   \centering
   \includegraphics[width=\textwidth, angle=0]{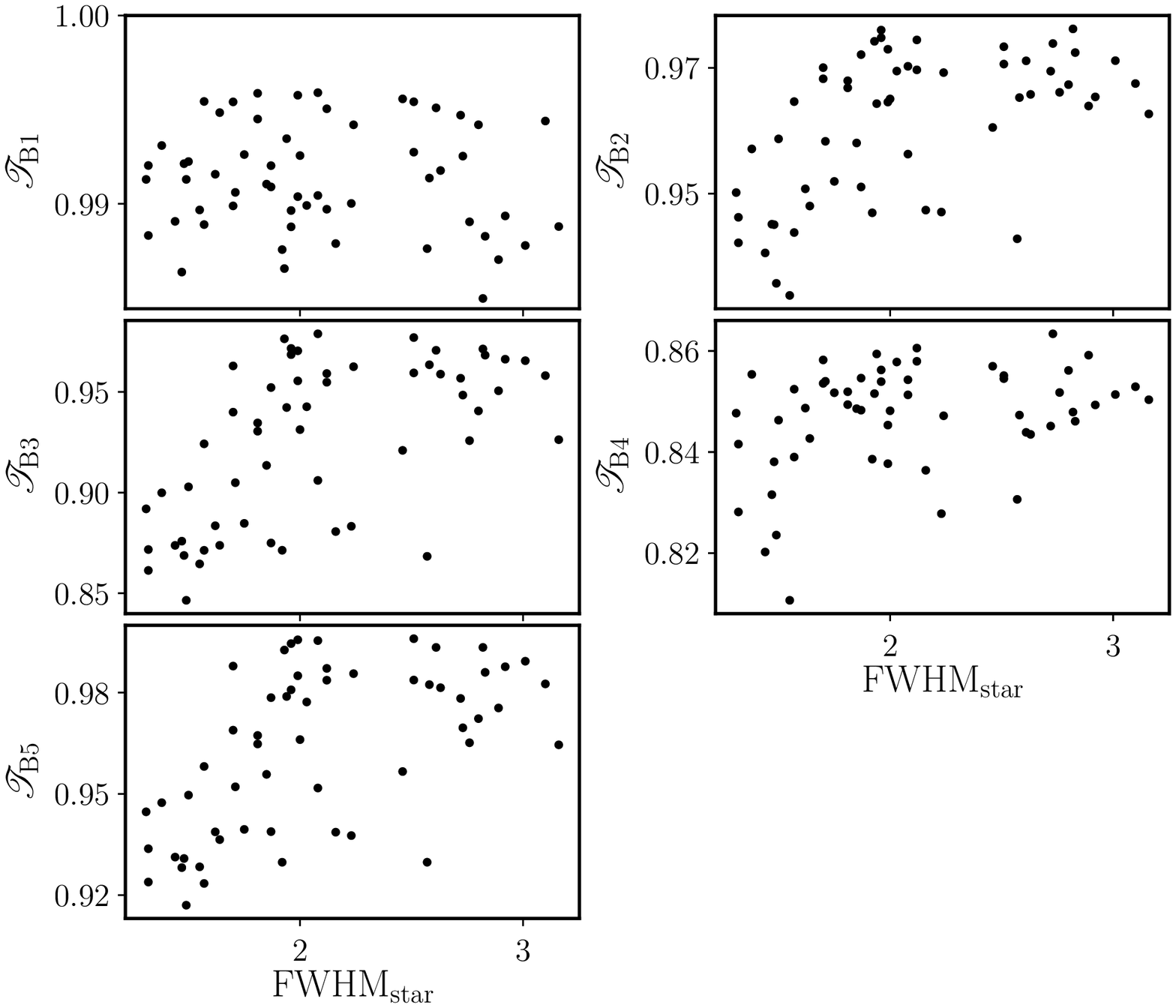}
   \caption{Relationship between the integral transparency and the width of star's flux distribution 
   ($\rm FWHM_{star}$, measured from the short exposure image observed before and near the spectroscopy). 
   The results of correlation analysis are listed in Table~\ref{Tab:sta}.}
   \label{Fig:relation2}
  \end{figure}

We calculated the airmass in the middle of exposure time, and investigated the relation between the integral transparency and airmass. 
The integral transparency as a function of airmass was presented in Figure~\ref{Fig:relation}. We calculated the Spearman correlation coefficient 
between the integral transparency and airmass and listed in Table~\ref{Tab:sta}. 
These analyses show that the integral transparency of the $\rm O_{2}$ absorption band is well correlated with airmass, 
and the integral transparency of the $\rm H_{2}O$ absorption bands also be effected by airmass. 
In addition, we measured the width ($\rm FWHM_{star}$) of star's flux distribution from the short exposure image observed before and near the spectroscopy, 
and showed the relation between the integral transparency and the $\rm FWHM_{star}$ in Figure~\ref{Fig:relation2}. 
We also calculated the Spearman correlation coefficient between the integral transparency and the $\rm FWHM_{star}$ and listed in Table~\ref{Tab:sta}. 
In practice, the variation of $\rm FWHM_{star}$ is mainly modulated by varying seeing, where seeing is always changing. 
These analyses show that the integral transparency of the $\rm H_{2}O$ absorption bands correlates with the variation of $\rm FWHM_{star}$. 

\begin{table}
\begin{center}
\caption[]{Statistics of the telluric absorption features. Col~(1) is the integral transparency of different absorption bands. Col~(2) is the absorber. 
The minimum, maximum and averaged value of the integral transparencies for the different absorption bands are listed in Col~(3),~(4) and~(5), respectively. 
Col~(6) is the correlation analysis results ($r$, $p$) between the integral transparency and airmass. 
Col~(7) is the correlation analysis results ($r$, $p$) between the integral transparency and the width of star's flux distribution.}\label{Tab:sta} 
 \begin{tabular}{lccccccc}
  \hline\noalign{\smallskip}
$\mathscr{T}_{\rm B*}$  &  Molecular &  Min($\mathscr{T}_{\rm B*}$) & Max($\mathscr{T}_{\rm B*}$) & $\widehat{\mathscr{T}_{\rm B*}}$   & Airmass       & $\rm FWHM_{star}$ \\
(1)&(2)&(3)&(4)&(5)&(6)&(7)\\
  \hline\noalign{\smallskip}
$\mathscr{T}_{\rm B1}$  & $\rm O_{2}~(\gamma~band)$         & 0.985 & 0.996 & 0.991$\pm$0.003 & (-0.25, 0.07)                           &(-0.07, 0.58)        \\
$\mathscr{T}_{\rm B2}$  & $\rm O_{2}~(B~band) + H_{2}O$    & 0.934 & 0.976 & 0.961$\pm$0.011 & (-0.52, 3.91$\times10^{-5}$)  & (0.53, 1.85$\times 10^{-5}$) \\
$\mathscr{T}_{\rm B3}$  & $\rm H_{2}O$                                  & 0.846 & 0.979 & 0.926$\pm$0.039 & (-0.41, 1.52$\times10^{-3}$)  & (0.60, 8.36$\times 10^{-7}$) \\
$\mathscr{T}_{\rm B4}$  & $\rm O_{2}~(A~band)$                    & 0.811 & 0.863 & 0.847$\pm$0.011 & (-0.90, 9.56$\times10^{-21}$) & (0.28, 0.03)                           \\
$\mathscr{T}_{\rm B5}$  & $\rm H_{2}O$                                  & 0.917 & 0.996 & 0.964$\pm$0.024 & (-0.43, 7.71$\times10^{-4}$)   & (0.59, 1.22$\times 10^{-6}$) \\
  \noalign{\smallskip}\hline
\end{tabular}
\end{center}
\end{table}

\section{Discussion}
\label{sec:dis}
In the ground-based optical spectrum, the telluric absorption bands are caused by two kinds of absorbers: 
molecular oxygen ($\rm O_{2}$) and water vapour ($\rm H_{2}O$). Based on the long slit spectroscopy, 
we provide a accurate correction method for the telluric absorption lines using the observed comparison 
star spectrum and stellar spectral library, the correction accuracy is 1\%. 
Our correction method for the telluric absorption lines has many advantages over the traditional method 
(i.e., observing the telluric standard star), especially in the study of long-term spectral monitoring. For example, 
(1) the scientific object and its comparison star are observed simultaneously in the long slit with the same observing conditions, 
such as the same airmass and seeing. Therefore, It is possible that the correction accuracy of the telluric absorption lines using 
this method (1\%) should be better than the traditional method; 
(2) in the long-term spectral monitoring project, we do not need to take expensive telescope time to observe the telluric standard star in each sampling. 
Therefore, in the research of one-epoch spectral observation or long-term spectroscopic monitoring, 
this work provides a kind of worthy observation strategy and corresponding method to correct the telluric absorptions. 

As described in Section~\ref{sec:method}, the telluric transmission spectrum is obtained 
from the process of fitting the observed spectrum of comparison star by stellar template through masking the discrete absorption windows. 
The discrete absorption windows mean we can select the continuum windows in the spectral fitting process. 
In this work, we only apply this method described in Section~\ref{sec:method} to the spectrum 
that its wavelength covers less than 9000~\AA, which is limited by the spectrograph of YFOSC. 
Actually, because the wavelength ranges from 9000~\AA~to around 17000~\AA~exist the available continuum windows 
(see the synthetic telluric absorption spectrum calculated by \cite{Smette2015} using the Line by Line Radiative Transfer Model, 
the telluric absorption bands are also discrete), the application of this method can be extrapolated from optical region to 17000~\AA. 

The telluric transparency of B4 absorption band caused by oxygen has the largest value, 
and has the strongest correlation with airmass but it  has no correlation with the seeing. 
While, the telluric transparency of B1 absorption band also caused by oxygen has the smallest value, 
which does not correlate with airmass and seeing (see Section~\ref{sec:lo}). 
However, not only the airmass but also seeing correlates with the telluric transparency of water vapour absorption (e.g., B3 and B5 bands). 
These results show that (1) the absorptions of oxygen and water vapour in the optical spectrum closely correlate with atmospheric depth, 
and (2) the oxygen concentration of observation site does not change with the varying seeing 
while the amount of water changes with the varying seeing. 
In addition, analysis results of Section~\ref{sec:lo} show that the red side of optical spectrum including emission lines and continuum 
would be affected by the varying telluric absorptions, accurate correction of the telluric absorption lines should benefit from spectral flux calibration. 

\section{Summary}
\label{sec:sum}
In this paper, combining the advantages of long slit spectroscopy, we described a method that can be use to accurately 
correct the telluric absorption lines of spectrum. We constructed the telluric transmission spectrum 
by fitting the observed spectrum of comparison star by the stellar template, 
then obtained the telluric absorption corrected spectrum by dividing the observed spectrum by the telluric transmission spectrum. 
We applied this method to correct the telluric absorption lines of the long-term monitored spectra  by Lijiang 2.4-meter telescope. 
Using the scatter of the [O~{\sc III}]~$\lambda$5007 fluxes emitted from the narrow-line region of active galactic nuclei (SDSS J153636.22+044127.0) 
as an indicator,  we found that the correction accuracy of this method for the telluric absorption lines is 1\%. 
We investigated the telluric absorption features of Lijiang Observatory, 
and found that the telluric absorption transparency correlates with the relative humidity, airmass and seeing. 
This work provides a kind of worthy observation strategy and corresponding method to correct the telluric absorptions. 

\begin{acknowledgements}
We are grateful to the referee for useful suggestions that improved the manuscript. 
This work is supported by the National Natural Science Foundation of China (NSFC; 11991051, 12073068, 11703077, and 11803087). 
K.X.L. acknowledges financial support from from the Yunnan Province Foundation (202001AT070069), 
and from the Light of West China Program provided by Chinese Academy of Sciences (Y7XB016001). 
L.X. acknowledges financial support from the Light of West China Program provided by Chinese Academy of Sciences (Y8XB018001). 
We acknowledge the support of the staff of the Lijiang 2.4 m telescope. 
Funding for the telescope has been provided by the CAS and the People’s Government of Yunnan Province. 
\end{acknowledgements}

\end{document}